\documentclass[twocolumn, prb]{revtex4}

\usepackage{graphicx}
\usepackage{dcolumn}
\usepackage{bm}

\begin{document}

\title{  Renormalization group method for weakly-coupled  
 quantum chains:  application to the spin one-half Heisenberg model}

\author {S. Moukouri}

\affiliation{ Michigan Center for Theoretical Physics and
             Department of Physics, \\
         University of Michigan 2477 Randall Laboratory, Ann Arbor MI 48109}

\begin{abstract}
The Kato-Bloch perturbation formalism is used to present a density-matrix
renormalization-group (DMRG) method for strongly anisotropic two-dimensional
systems. This method is used to study Heisenberg chains 
weakly coupled by the transverse couplings $J_{ \perp}$ and $J_{d}$ 
( along the diagonals). An extensive comparison of the renormalization
group and quantum Monte Carlo results for parameters where the
simulations by the latter method are possible shows a very good
agreement between the two methods. It is found, by analyzing ground state 
energies and spin-spin correlation functions, that there is a transition
between two ordered magnetic states. When $J_{d}/J_{\perp} \alt 0.5$, 
the ground state displays a N\'eel order. When 
$J_{d}/J_{ \perp} \agt 0.5$, a collinear magnetic ground state
in which interchain spin correlations are ferromagnetic becomes stable.
In the vicinity of the transition point, $J_{d}/J_{ \perp} \approx 0.5$,
the ground state is disordered. But, the nature of this disordered
ground state is unclear. While the numerical data seem to show that the
chains are disconnected, the possibility of a genuine disordered 
two-dimensional state, hidden by finite size effects, cannot be excluded.
\end{abstract}

\maketitle

\section{Introduction}
\label{intro}

In a recent publication \cite{TS1-moukouri}, it was shown that the 
density-matrix renormalization group method (DMRG) \cite{white, DMRG-book} can
be applied to an array of weakly coupled quantum chains.
As an illustration of the method, weakly coupled Heisenberg spin chains
were studied and some partial results on the ground state energies
were shown to be in good agreement with previous quantum Monte Carlo
(QMC) studies.  But the essential question concerning the stability
of the disordered one-dimensional (1D) ground state against small
transverse perturbations was not addressed.

The motivation behind such
a study is in the search of a disordered ground state for a spin one-half
Heisenberg model in dimension higher than one.  A spin liquid state without 
spin rotational or translational symmetry breaking has been conjectured
to be relevant for the physics of high-Tc cuprate superconductors
 \cite{anderson1}. A possible candidate is the resonance valence bond (RVB) 
state \cite{anderson}. Earlier attempts \cite{chandra1,dagotto,chandra2,sorella}
to find the RVB ground state by various techniques ($1/S$ expansions, exact
diagonalization, quantum Monte Carlo) have given some indication about its
possible realization. But their conclusions are still disputed.
 It has even been argued \cite{read} that the spin-Peierls mechanism, not RVB,
may be the most natural way to lead to a disordered state.

\begin{figure}
\includegraphics[width=3. in]{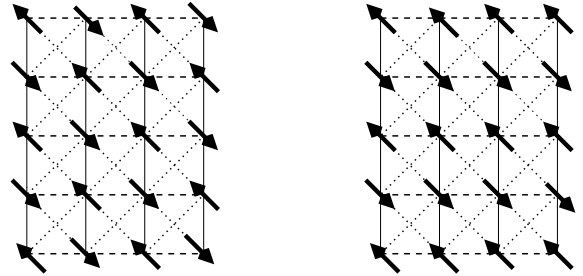}
\caption{Sketch of the ground state of weakly coupled Heisenberg chains as
function of $J_{ \perp}$ (dashed lines along the rungs) and $J_{d}$
(dotted lines along the diagonals): N\'eel state when 
$J_{d}/J_{ \perp} \alt 0.5$ (left), collinear state when 
$J_{d}/J_{ \perp} \agt 0.5$ (right)}
\label{gstate}
\end{figure}

More recently, the interest has shifted to search for a RVB state on
quasi 1D systems. A pure spin one-half Heisenberg chain has a disordered
ground state with neutral spin one-half excitations (spinons) and does
not break spin rotational or translational symmetry. It is thus tempting
to try to find a higher dimensional generalization of this state by
the application of small perturbations. Contrary to an earlier claim of the
realization of a spin liquid \cite{parola}, subsequent studies \cite{affleck,
rosner,sandvik2}  indicate
that the introduction of the rung transverse coupling $J_{ \perp}$ between 
the chains (see Fig.~\ref{gstate}) seems  to lead to a N\'eel state for any
non-zero $J_{ \perp}$. A possible way to avoid the N\'eel order is to
introduce, in addition to $J_{ \perp}$, a small frustration $J_{d}$ 
along the diagonals. In a recent work \cite{tsvelik} it was claimed that 
a spin liquid state is realized when $J_{ \perp}=2J_{d}$.

 A more direct motivation in studying a model of weakly coupled Heisenberg 
chains stems to its  relevance to the understanding of interchain effects in
quasi-one-dimensional materials \cite{sajita,keren,tennant}. A recent
neutron scattering experiment \cite{coldea} on the frustrated antiferromagnet
(AFM) $Cs_2CuCl_4$ found that the dynamical correlation show a highly 
dispersive continuum of a excitations with fractional quantum numbers,
 a signature of a spin liquid state. 

In this paper, a general formalism of the DMRG algorithm for weakly
coupled chains of Ref.\cite{TS1-moukouri} is presented.   
This method is a particular case of a recent matrix version \cite{MAT-moukouri} 
 of the general perturbation expansion 
which was proposed decades ago by Kato and Bloch \cite{kato,bloch,messiah}.
The Kato-Bloch expansion was initially introduced to find the correction
on a single state. This expansion is straightforwardly generalized to
account for many low lying states. The method is in spirit close to
an earlier perturbative renormalization group by Hirsch and Mazenko
\cite{hirsch}. A more detailed study of weakly coupled Heisenberg 
chains is presented. An extensive comparison with quantum Monte Carlo
results for unfrustrated transverse couplings is made. It shows a  
good agreement between the two methods when the perturbation is small and the
lattice not too large. Then the question of the stability
of the non-magnetic state in the presence of frustration is addressed. 
 It is shown that the 2D DMRG algorithm can provide
a convincing answer to this question, at least for the parameters
that were investigated. It is found, by analyzing ground state
energies and spin-spin correlation functions, that the perturbation is 
relevant, leading to magnetic ground states (see Fig.~\ref{gstate}). 
 When $J_{d}/J_{ \perp} \alt 0.5$, the ground state displays a 
N\'eel order. When $J_{d}/J_{ \perp} \agt 0.5$ a collinear magnetic 
ground state in which interchain spin correlations are ferromagnetic becomes 
stable. In the vicinity of the transition point,  
$J_{d}/J_{ \perp} \approx 0.5$, 
the system seems to behave as an assembly of independent chains.
This is reminescent of the so-called sliding Luttinger liquid \cite{lubensky}
 recently found in a model of crossed spin one-half Heisenberg chains
\cite{singh}. But it is impossible to exclude a genuine 2D spin liquid 
state (i.e., with a spin gap) masked by  finite size effects.

\section{Formal development}
\label{formal}

The DMRG method described below can work for spin, fermionic
as well as bosonic systems, and so it is convenient to use a general formulation
of the algorithm that can then  be adapted to each of these cases.
The model Hamiltonians under consideration can be written as follows:

\begin{eqnarray}
H=H_{\parallel}+gH_{\perp},
\label{hamilton}
\end{eqnarray}

\noindent  where $H_{\parallel}$ is a the sum over
one-dimensional (1D) Hamiltonians (longitudinal direction),

\begin{eqnarray}
H_{\parallel}= \sum_{l=1}^L H_l,
\label{hamilton2}
\end{eqnarray}

\noindent and $H_{\perp}$ is
the interaction between these 1D systems (transverse direction). The coupling 
constant $g$ is such that $g \ll 1$.

Since $g \ll 1$, it is natural to study the problem using perturbation
theory. The Kato-Bloch formalism is convenient to set up a perturbation 
expansion around a numerical solution of $H_{\parallel}$ provided by the DMRG.
 For a single chain $l$ whose Hamiltonian is $H_l$, a set of
 eigenstates $|\phi_{n_l}\rangle$ and eigenvalues  $\epsilon_{n_l}$
can be obtained by the usual 1D DMRG. The zeroth order set of eigenstates 
$|\Phi_{\parallel[n]}\rangle$ of the full longitudinal Hamiltonian is simply 
given by the tensor product of the  $|\phi_{n_l}\rangle$,

\begin{eqnarray}
|\Phi_{\parallel[n]}\rangle =  |\phi_{n_1}\rangle  |\phi_{n_2}\rangle ...  
|\phi_{n_L} \rangle,
\label{wavefunction}
\end{eqnarray}

\noindent and the set of approximate eigenvalues of  $H_{\parallel}$
 is given by the sum

\begin{eqnarray}
E_{\parallel[n]} =  \epsilon_{n_1}+  \epsilon_{n_2}+ ...  +\epsilon_{n_L},
\label{eigenvalue}
\end{eqnarray}

\noindent where $[n]=(n_1,n_2,...,n_L)$ and  $n_{l}$ labels to 
an eigenset on the chain $l$.

Let $P_{\parallel}$ be the projector on the states $|\Phi_{\parallel[n]}\rangle$,

\begin{eqnarray}
P_{\parallel} = \sum_{[n]} |\Phi_{\parallel[n]}\rangle 
\langle\Phi_{\parallel[n]}| 
\label{projector}
\end{eqnarray}  

\noindent and $Q_{\parallel}=1-P_{\parallel}$. 

 Let ($E_{[n]}$, $|\Phi_{[n]}\rangle$) be the exact eigenset of $H$. This 
eigenset will tend to ($E_{\parallel[n]}$, $|\Phi_{\parallel[n]}\rangle$)
in the limit $g \rightarrow 0$. Let $P$ be the projector onto the
spaces  $|\Phi_{[n]}\rangle$. $P$ may be written as follows

\begin{eqnarray}
P = \sum_{[n]} |\Phi_{[n]}\rangle \langle\Phi_{[n]}|.
\label{projector2}
\end{eqnarray}

Since the perturbation $g$ is small, it is assumed that the subspaces 
generated by the $|\Phi_{\parallel[n]}\rangle$'s and by the 
$|\Phi_{[n]}\rangle$'s are 
not orthogonal. An approximate expression of $HP$ in the basis spanned by 
the eigenstates of $H_{\parallel}$ will now be derived by using a 
 generalization of a method first introduced by Kato \cite{kato} and
later modified by Bloch \cite{bloch}.  The advantage of the Bloch's version
is that it leads to a simpler expansion.

 Following Bloch, let ${\cal U}$ be the operator

\begin{eqnarray}
{\cal U} = \sum_{[n]} |\Phi_{[n]}\rangle \langle\Phi_{\parallel[n]}|
\label{defu} 
\end{eqnarray}

which projects the  $|\Phi_{\parallel[n]}\rangle$ onto $|\Phi_{[n]}\rangle$, 
 and ${\cal U}$ satifies, ${\cal U} P_{\parallel} = {\cal U}$. The problem 
of finding an expansion of $HP$ projected onto $P_{\parallel}$
is equivalent to finding an expansion for $P_{\parallel} H {\cal U}$.

One starts by deriving an equation satisfied by ${\cal U}$. 
 The Schr\"odinger equation

\begin{eqnarray}
H|\Phi_{[n]}\rangle=E_{[n]}|\Phi_{[n]}\rangle
\label{shrodinger}
\end{eqnarray}

\noindent is transformed as follows,

\begin{eqnarray}
(H-{\tilde H}_{\parallel})|\Phi_{[n]}\rangle= {\cal E}_{[n]} |\Phi_{[n]}\rangle
\label{schrodinger2}
\end{eqnarray}

\noindent where ${\tilde H}_{\parallel}$ is identical to $H_{\parallel}$ in the subspace
spanned by  the $|\Phi_{\parallel[n]}\rangle$'s, and, 

\begin{eqnarray}
{\cal E}_{[n]}= E_{[n]}-\langle\Phi_{[n]}|{\tilde H}_{\parallel}
|\Phi_{[n]}\rangle.
\label{corrE} 
\end{eqnarray}

When a single state $|\Phi_{\parallel[0]}\rangle$ is kept, 
${\tilde H}_{\parallel}$ is given by

\begin{eqnarray}
\nonumber {\tilde H}_{\parallel} = \left( 
\begin{array}{ccccccc} 
E_{\parallel [0]} & 0 & 0 & 0 & 0 & \ldots & 0 \\
0 & E_{\parallel [0]} & 0 & 0 & 0 & \ldots & 0 \\
0 & 0 & E_{\parallel [0]} & 0 & 0 & \ldots & 0 \\
  &   & \ldots &  &  & \ldots &   \\
0 & 0 & \ldots & E_{\parallel [0]} & &  \ldots &   \\
0 & 0 & 0 & \ldots & E_{\parallel [0]} & \ldots & 0  \\
  &   & \ldots &  &  & \ldots &   \\
0 & 0 & 0 & 0 & 0 & \ldots & E_{\parallel [0]} \\
\end{array}
\right).
\label{block1}
\end{eqnarray}

\noindent The method reduces to the usual stationary perturbation
expansion. It is known that such an expansion does not often converge.
 The main source of divergence is the near degeneracy of the eigenvalues.
Now if many states up to a cut-off $n_c$  are kept, a possible generalization of
${\tilde H}_{\parallel}$ to many states 
$|\Phi_{\parallel [0]}> \ldots |\Phi_{\parallel [n_c]>}$, is

\begin{eqnarray}
\nonumber {\bar H}_{\parallel} = \left( 
\begin{array}{ccccccc} 
E_{\parallel [0]} & 0 & 0 & 0 & 0 & \ldots & 0 \\
0 & E_{\parallel [1]} & 0 & 0 & 0 & \ldots & 0 \\
0 & 0 & E_{\parallel [2]} & 0 & 0 & \ldots & 0 \\
  &   & \ldots &  &  & \ldots &   \\
0 & 0 & \ldots & E_{\parallel [n_c]} & &  \ldots &   \\
0 & 0 & 0 & \ldots & E_{\parallel [0]} & \ldots & 0  \\
  &   & \ldots &  &  & \ldots &   \\
0 & 0 & 0 & 0 & 0 & \ldots & E_{\parallel [0]} \\
\end{array}
\right).
\label{block2}
\end{eqnarray}

\noindent Thus if $n_c$ is suitably chosen, the
series will converge \cite{MAT-moukouri}. The purpose of this choice is 
to shield the eigenvalue $E_{\parallel [0]}$ from the rest of the spectrum 
by treating the $n_c-1$ states just above the ground state exactly and 
the remaining spectrum perturbatively.


By applying 
$P_{\parallel}$ and then ${\cal U}$ to the  Equation(~\ref{schrodinger2}) 
above, one finds,

\begin{eqnarray}
g{\cal U} H_{\perp} |\Phi_{[n]}\rangle= {\cal E}_{[n]} |\Phi_{[n]}\rangle
\label{schrodinger3}
\end{eqnarray}

The subtraction of Equation(~\ref{schrodinger3}) from
Equation(~\ref{schrodinger2}) leads to

\begin{eqnarray}
(H-{\tilde H}_{\parallel}-g{\cal U} H_{\perp})|\Phi_{[n]}\rangle = 0. 
\label{schrodinger4}
\end{eqnarray}

 By applying $\langle\Phi_{\parallel [n]}|$ on the right of 
equation(~\ref{schrodinger4}) and performing the summation  over $[n]$ , 
one finally obtains the equation satisfied by ${\cal U}$,

\begin{eqnarray}
(H-{\tilde H}_{\parallel}-g{\cal U} H_{\perp}){\cal U} = 0. 
\label{schrodinger5}
\end{eqnarray}

 Equation(~\ref{schrodinger5}) is further transformed by
using the fact that $P_{\parallel} {\cal U} = P_{\parallel}$ and 
${\cal U} = P_{\parallel}{\cal U} + Q_{\parallel} {\cal U}$. One obtains: 

\begin{eqnarray}
{\cal U}=P_{\parallel} + g{\tilde Q}_{\parallel}(H_{\perp}{\cal U}-
        {\cal U} H_{\perp} {\cal U}) 
\end{eqnarray}

\noindent where  $\tilde{Q_{\parallel}}$ is given by

\begin{eqnarray}
\tilde{Q_{\parallel}}= Q_{\parallel} 
({\tilde H}_{\parallel}-H_{\parallel})^{-1}. 
\label{projector3}
\end{eqnarray}

 This leads to the expansion for ${\cal U}$

\begin{eqnarray}
 {\cal U}^{(0)}=P_{\parallel} 
\label{expansion0}
\end{eqnarray}
\begin{eqnarray}
{\cal U}^{(n)}= g{\tilde Q}_{\parallel}[H_{\perp}{\cal U}^{(n-1)}-
               \sum_{p=1}^{n-1}{\cal U}^{(p)} H_{\perp}{\cal U}^{(n-p-1)}]
\label{expansion1}
\end{eqnarray}

From this expansion, one finds the approximate 
Hamiltonian ${\tilde H}=P_{\parallel} H {\cal U}$ is 

\begin{eqnarray}
\nonumber \tilde{H}= \sum_{[n]} E_{\parallel [n]} |\Phi_{\parallel [n]}\rangle 
\langle\Phi_{\parallel [n]}| + 
 g P_{\parallel} H_{\perp}P_{\parallel}+ \\ 
g^2 P_{\parallel} H_{\perp}\tilde{Q_{\parallel}} H_{\perp}P_{\parallel}+... 
\label{expansion2}
\end{eqnarray}

This perturbation expansion is a matrix generalization to
 many states of Bloch's expansion \cite{bloch,messiah} which was 
established for a
single state. Even though the ground state and a few low lying states
will ultimately be computed, it is important to keep many low
lying states in the perturbation expansion. This is because the 
convergence will mainly depend on two quantities. The
first one is obviously $g$. The second one is the
projector $\tilde{Q_{\parallel}}$. 
If ${\tilde H}_{\parallel}={H}_{\parallel}$, then  $Q_{\parallel}=0$.  
 In that case only the first order term
in equation (~\ref{expansion2}) is not equal to zero. The rewriting
of the original problem to equation  (~\ref{expansion2}) is a simple
change of basis. So in the limit where $n_c=dim A$, where $A$ is the
Hilbert space in which all the operators are defined, the method
is exact. But since only a small number of eigenstates of 
$H_{\parallel}$ can be used even if the full spectrum is known, 
$Q_{\parallel} \neq 0$. The magnitude of  $\tilde{Q_{\parallel}}$ in the
expansion decreases by increasing the cut-off $n_c$.
 It is to be noted this matrix expansion is close to
the method of Hirsch and Mazenko \cite{hirsch}, who also
used a block expansion near the solution of an unperturbed
Hamiltonian. The problem with their study was, however, that their 
technique was applied to a model with no small parameter.

 When the DMRG is used as a method of solution for $H_{\parallel}$,
 we can not know  $Q_{\parallel}$ exactly.  This is 
because the DMRG does not keep any information about the
truncated states. But it is possible to define a perturbative expansion
in a reduced space spanned by the states kept. The above perturbative 
expansion will thus be
adapted in this study as follows. During the 1D DMRG part of
the method, $N_s=ms_1 \times ms_1$ states will be obtained for the reduced
superblock (i.e., the superblock reduced to the two external
large blocks; it is supposed that open boundary conditions (OBC) 
are used). Typically $ms_1=16-192$ during this investigation.
 The complete spectrum of 
this reduced superblock can be obtained as in the thermodynamic
algorithm \cite{moukouri-THERMO}. This spectrum will serve as
 $H_{\parallel}$. Only
a small fraction $ms_2=16-96$ of these states can be kept for the generation
of the 2D lattice. The $ms_2$ states will define $P_{\parallel}$, and   
 $Q_{\parallel}$ is constructed using the
remaining $Ns-ms_2$ states. Hence the perturbation expansion in 
Eq.(~\ref{expansion2}) will be made by assuming that $H_{\parallel}$
is the low energy Hamiltonian of size $ms_1 \times ms_1$ obtained
from the DMRG rather than the exact 1D solution of $H_{\parallel}$.

 The Hamiltonian $ \tilde{H}$ is one-dimensional and it will be studied by the
DMRG method. The only difference with a normal 1D situation
is that the local operators are now $ms_2 \times ms_2$ matrices
which makes computations heavier. It should be noted that the accuracy of
the method is related to the diagonalized unperturbed Hamiltonian obtained 
from the DMRG. This Hamiltonian, although it leads to a very accurate
ground state energy, is less accurate for high lying states and correlation
functions. So the potential errors of the method will come from the DMRG
as well as the truncated perturbative series. A better approach is to
use the exact diagonalization method to diagonalize the unperturbed Hamiltonian.
 However, in that case one will be restricted to small chains.

\section{Application to the Heisenberg model}
\label{application}

 The above formalism will now be applied to the anisotropic Heisenberg
model on a 2D square lattice. The Hamiltonian reads:

\begin{eqnarray}
 \nonumber H_{spins}=\sum_{i,l}{\bf S}_{i,l}{\bf S}_{i+1,l}+J_{ \perp}
\sum_{i,l}{\bf S}_{i,l}{\bf S}_{i,l+1}+ \\
J_{d} \sum_{i,l}({\bf S}_{i,l}{\bf S}_{i+1,l+1}+
{\bf S}_{i+1,l}{\bf S}_{i,l+1})
\label{heisenberg}
\end{eqnarray}

\noindent where the ${\bf S}_{i,l}$ are the usual spin one-half operators.

The question of the condition of the onset of long-range order as a function
of $J_{\perp}$ has been addressed in many studies. Spin-wave analysis
\cite{sakai,parola} predicted that there is a finite critical 
$J_{ \perp c} \approx 0.03$,  
above which long-range order is established. Renormalization
group analysis \cite{affleck} supplemented by series expansion
 computations found that if $J_{ \perp c}$ is finite, it cannot exceeds $0.02$.
A finite critical value is at variance with a random phase approximation
 (RPA) \cite{rosner} which predits $J_{ \perp c}=0$.  The QMC method combined
 with a multichain mean-field approach \cite{sandvik2} has concluded
 that when $J_{ \perp}=0$, the ground state is an antiferromagnet down
to $J_{ \perp} =0.02$. From these studies, it is likely that the AFM 
ground state is stable as soon as $ J_{ \perp} \neq 0$. This does not, 
however, preclude a spin liquid ground state in the case when 
$J_{d}$ is added
between the chains. When this exchange term is added, the QMC method
faces the infamous sign problem. The two-step DMRG method presented
here can help to find, if it exists, the spin liquid ground state.

The adaptation of the formalism discussed in section(~\ref{formal})
to the model of Equation(~\ref{heisenberg}) is without any difficulty. 
The first step is the solution of the 1D Hamiltonian:

\begin{eqnarray}
H_{l}=\sum_{i}{\bf S}_{i,l}{\bf S}_{i+1,l}
\label{chain}
\end{eqnarray}

\noindent by the usual DMRG method. This yields the chain eigenvalues 
$\epsilon_{n_l}$ and eigenstates
$|\phi_{n_l}\rangle$. From equation (~\ref{expansion2}), the projected
Hamiltonian in the first order approximation is given by

\begin{eqnarray}
\nonumber \tilde{H}= \sum_{[n]} E_{\parallel [n]} |\Phi_{\parallel [n]}\rangle 
\langle\Phi_{\parallel [n]}| + 
J_{ \perp} P_{\parallel} \sum_{i,l}{\bf S}_{i,l}{\bf S}_{i,l+1} 
P_{\parallel}+ \\ 
J_{d} P_{\parallel} \sum_{i,l}({\bf S}_{i,l}{\bf S}_{i+1,l+1}+ 
{\bf S}_{i+1,l}{\bf S}_{i,l+1}) P_{\parallel}.
\label{expansion3}
\end{eqnarray}

\noindent which may simply be written as

\begin{eqnarray}
 \nonumber \tilde{H} \approx \sum_{[n]} E_{\parallel [n]} |\Phi_{\parallel [n]}
\rangle \langle\Phi_{\parallel [n]}| +
 J_{ \perp} \sum_{il} {\bf \tilde{S}}_{i,l} {\bf \tilde{S}}_{i,l+1}+ \\ 
 J_{d} \sum_{il} ({\bf \tilde{S}}_{i,l} {\bf \tilde{S}}_{i+1,l+1}+
{\bf \tilde{S}}_{i+1,l} {\bf \tilde{S}}_{i,l+1}),
\end{eqnarray}

\noindent where ${\bf {\tilde S}}_{i,l}^{n_l,m_l}=\langle \phi_{n_l}
|{\bf S}_{i,l}|\phi_{m_l}\rangle$.

The matrix elements for the second
terms may be written

\begin{widetext}

\begin{eqnarray}
\langle \Phi_{\parallel[n]}|H_{\perp}{\tilde Q}_{\parallel}H_{\perp}
|\Phi_{\parallel[n']}\rangle=  
\sum_{il,i'l',[m]} \frac{\langle\Phi_{\parallel[n]}| {\bf S}_{i,l}
{\bf S}_{i,l+1}|\Phi_{\parallel[m]}\rangle 
\langle \Phi_{\parallel[m]}| {\bf S}_{i,l}
{\bf S}_{i,l+1}|\Phi_{\parallel[n']}\rangle}{E_{\parallel [0]}-E_{\parallel [m]}}
\label{mat2}
\end{eqnarray}

\end{widetext}

 The second order term (~\ref{mat2}) generates a long-range coupling 
between the chains, which makes it difficult to treat.
One can see that the condition for the matrix element to be non zero is
that $\langle\phi_{n_{l1}}|\phi_{m_{l2}}\rangle=\delta_{n_{l1},m_{l2}}$ 
except when $l_{1,2}=l$ or $l+1$. Thus,

\begin{widetext}

\begin{eqnarray}
E_{[m]}-E_{[0]}=(\epsilon_{n_1}-\epsilon_{0_1})+...+(\epsilon_{m_l}-
\epsilon_{0_l})+
(\epsilon_{m_{l+1}}-\epsilon_{0_{l+1}})+...+(\epsilon_{n_L}-\epsilon_{0_L}).
\label{ediff}
\end{eqnarray}

\end{widetext}

In the Eq.~\ref{ediff} above, the dominant terms will come from the differences
involving the indices $m_l$ and $m_{l+1}$ because the others terms come from
the state used to generate $P_{\parallel}$ and are thus of lower energies.
Up to the second order, the effective one-dimensional Hamiltonian, which
is written here without the frustration term, is

\begin{eqnarray}
\nonumber \tilde{H} \approx \sum_{[n]} E_{\parallel [n]} 
|\Phi_{\parallel[n]}\rangle \langle \Phi_{\parallel [n]}| +
 J_{\perp} \sum_{l} {\bf \tilde{S}}_{l} {\bf \tilde{S}}_{l+1}-\\
\frac{J_{\perp}^2}{2} \sum_{l} {\bf S}_{l}^{(2)}{\bf S}_{l+1}^{(2)}
+...
\label{efhamil}
\end{eqnarray}

\noindent where the chain-spin operators on the chain $l$ are
${\bf \tilde{S}}_{l}=({\bf \tilde{S}}_{1l},
 {\bf \tilde{S}}_{2l}, ...{\bf \tilde{S}}_{Ll})$ and
${\bf S}_{il}^{(2)}=({\bf \tilde{S}}_{1l}^{(2)},
 {\bf \tilde{S}}_{2l}^{(2)}, ...{\bf \tilde{S}}_{Ll}^{(2)})$, $L$ is the
chain length. The matrix elements of the  second order local
spin operators are 

\begin{eqnarray}
{\bf S}_{il}^{(2)n_ln'_l}=\sum_{m_l}\frac{{\tilde S}_{il}^{n_lm_l}
{\tilde S}_{il}^{m_ln'_l}}{\sqrt{\epsilon_{m_l}-\epsilon_{0_l}}}.
\end{eqnarray}

\noindent One can note that this expression of ${\bf S}_{il}^{(2)}$ is not
exact, it has been simplified to avoid long-range coupling between the
chains. The effective 1D Hamiltonian $\tilde{H}$ is also studied
using DMRG.

\section{Algorithmic details}
\label{sectalgo}

The algorithm of the method will now be described below. It 
consists of two DMRG steps separated by an intermediate stage in
which a simple block decimation is made.

\subsection{Step 1}

\begin{figure}
\includegraphics[width=3. in]{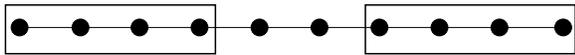}
\caption{sketch of the superblock in the step 1}
\label{step1}
\end{figure}

The first step of the method is the usual DMRG method for a single
chain.  The chain is divided into four blocks, and the two internal blocks
are made of a single site each. In the calculations, $ms_1=16-192$ 
states  are kept in the two external blocks.
 In most cases, the initial iteration starts with a chain having the
largest size before truncation, for instance $L=16$ when $ms_1=128$
states are kept. This way,
 a high accuracy is obtained even when the infinite system method is
used. During this step, the local spin operators ${\bf S}_{i}$ on
each site $i$ of the chain are stored and longitudinal spin-spin correlations
 $C(i,r)=\langle{\bf S}_{i}{\bf S}_{i+r}\rangle$ are also computed and stored.
 As discussed by Caron and Bourbonnais \cite{caron}, open boundary 
conditions (OBC) which are used here introduce spurious behavior at the 
edges of the chain. It is therefore better to chose the origin $i$ in 
$C(i,r)$ in the middle of chain. It is crucial during this step to target 
more just than the $S_z=0$ sector in order to
obtain a correct low-energy Hamiltonian. In addition to $S_z=0$, 
$S_z= \pm 1$, $\pm 2$ were targeted in this study.

\subsection{Block transformation}

\begin{figure}
\includegraphics[width=3. in]{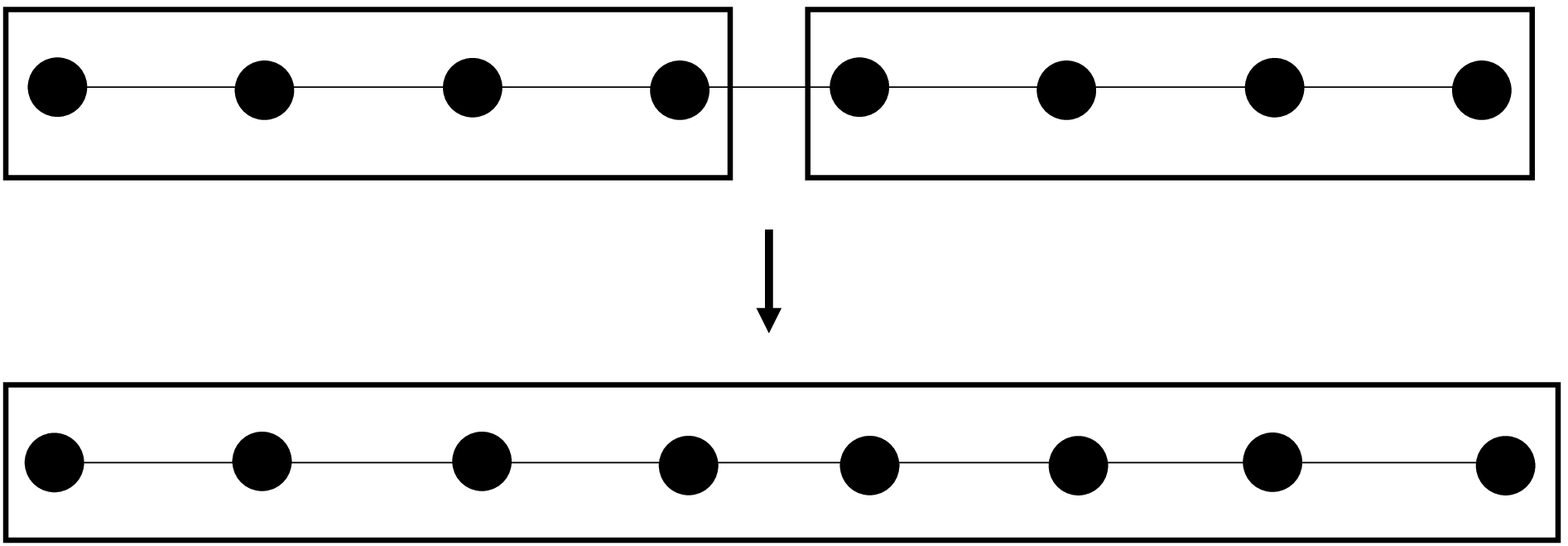}
\caption{sketch of the transformation of the two external blocks of 
length $L/2$ into a single block of length $L$ which is used as the 
building unit in step 2}
\label{step_int}
\end{figure}

An intermediate stage of the algorithm is a decimation process as in the old
block renormalization group method \cite{weinstein,jullien}.  In this
process, the two external blocks having $L/2$ sites each are reduced to
a single block with $L$ sites. During
this step, the $ms_1 \times ms_1$ states describing the chain are reduced
to $ms_2$ lowest states of the chain. As noted in Ref. \cite{TS1-moukouri},
since the block transformation is used only one time, the problem
of the propagation of spurious boundary effects \cite{white-noack} 
is not present. All the local spin operators and spin-spin correlation
functions are expressed in the basis of the $ms_2$ states.

\subsection{Step 2}

\begin{figure}
\includegraphics[width=3. in]{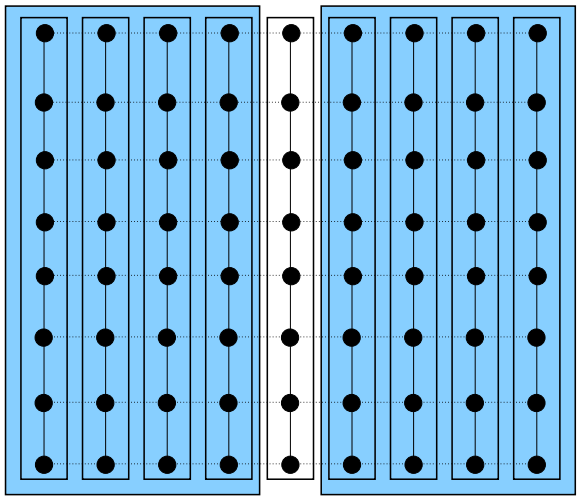}
\caption{ Sketch of the superblock in step 2}
\label{step2}
\end{figure}

The second step consists of applying the 1D DMRG method using the chains
obtained at the end of the previous step as the building blocks. This
step is indeed identical to the first step, except for the dimension of the
local spin operators. The central block is the chain from the previous step 
and thus has the dimension $ms_2 \times ms_2$. Typically, $ms_2=16-96$ and for 
the two external blocks, roughly the same number of states is kept.
If four blocks were taken as in the first step, the dimension of the 
superblock would be $ms_2^4$ which can become rapidly impratical.
 To ease the computations, three blocks instead of four are mostly used
during this step. As will be seen below, it is important during this step
to check that, for a given value of the couplings $J_{ \perp}$ and
$J_{d}$, enough states are kept such that a valid computation
is made. i.e.,  that the truncated Hamiltonian generated for the 
single chain is accurate enough, for the ground state and for the 
low lying states, to be used as a building block for the 2D lattice. One can
easily see that for a fixed $L$ and $J_{\perp,d} \ll
\Delta_{\sigma}(L)$, $\Delta_{\sigma}(L)$ is the finite size spin gap, and
the interchain matrix elements will be
negligible. The system will behave as a collection of free chains
even if $J_{\perp,d}$ is turned on. Now if $J_{\perp,d} \sim \delta
E(L)$, where $\delta E(L)$ is the width of the retained states,
the matrix elements of the states having higher energy,
which have been truncated out, have a non-negligible contribution.

\subsection{Algorithm}

The algorithm is summarized below.
\begin{itemize}

\item{
1. Build the low energy Hamiltonian for a single chain $B \cdot \cdot B$ 
       by using the 1D DMRG algorithm of Ref\cite{white}. 
   Store the spin operator ${\bf S}_{i}$ on each site and the correlation 
  function $C(i,r)$. 

}
\item{

2. When the block $B$ size is $L/2$, apply the block method 
    to merge the two external blocks into a single  
   block defined by the $ms_2$ states kept.  
   Express all the spin operators and correlation functions in 
   the basis of the $ms_2$ states. Check if for the number of states
   kept, the transverse couplings satisfy 
   $\Delta_{\sigma}(L) \alt J_{\perp,d} \ll \delta E(L)$. If this 
    condition is not satisfied, increase $ms_2$.  

}
\item{

3. Start a second 1D DMRG simulation identical to the first one 
   except that the central block is now a 
   single chain instead of a site, and the exchange coupling 
   is $J_{\perp,d}$ instead of $1.0$.
}
\end{itemize}

\section{ Results with four blocks in step 2}

In this part, the DMRG results are compared to the stochastic
series expansion (SSE) QMC results. The SSE-QMC method \cite{sandvik1}
is so far the most reliable technique for the study of 
quantum spin systems. It has been used to study weakly coupled 
quantum spin chains \cite{sandvik2}. It will thus be very
 instructive to see how well the DMRG method compares to the SSE-QMC.

\subsection{First order ground-state energies}

In Table (~\ref{table1}), the ground state energy per site for 
$12 \times 12$ systems
for  $m=16$, $24$ and $32$ is shown. In this calculation, four blocks were
used in the second DMRG step. The agreement with the SSE-QMC results is
 good for small $g$. The DMRG energies are higher than those of the QMC for all
transverse couplings studied. As $ms_2$ is increased, the difference 
between the DMRG and QMC energies decreases. This was expected since
the current method as the original DMRG procedure is variational.

\begin{table}
\begin{ruledtabular}
\begin{tabular}{ccccc}
 $J_{ \perp}$ & $ms_2=16$ & $ms_2=24$ & $ms_2=32$ & QMC  \\
\hline
 $0.00$ & -0.42848 & -0.42851 & -0.42851 & -0.42849(2) \\
 $0.05$ & -0.42900 & -0.42907 & -0.42909 & -0.42926(2) \\
 $0.10$ & -0.43058 & -0.43078 & -0.43090 & -0.43147(2) \\
 $0.15$ & -0.43312 & -0.43361 & -0.43387 & -0.43530(2) \\
 $0.20$ & -0.43642 & -0.44733 & -0.43780 & -0.44064(2) \\
 $0.25$ & -0.44028 & -0.44174 & -0.44247 & -0.44727(2) \\
\end{tabular}
\end{ruledtabular}
\caption{ DMRG ground state energies for $12 \times 12 $ lattices for 
$ms_1=ms_2=16$, $24$, and $32$ versus QMC.}
\label{table1}
\end{table}

The band-width of the states kept 
is $\Delta E = 1.132$, $1.290$, $1.466$ when $ms_2=16$, $24$ and
$32$ respectively. The target states during the first DMRG step were
  the lowest states of the spin sectors with $S_z=0$, $\pm1$ $\pm2$. The 
lowest states of higher spin sectors have energies which are higher than
the highest state kept in lower spin sectors, therefore they were not
targeted. The fact that the DMRG results compare well with the QMC ones
even at intermediate couplings reveals that for the spin chain, reliable
calculations can be made for values of $\delta E(L)/J_{ \perp} \approx 5$.
 But as expected for higher values of $J_{ \perp}$, the condition
 $\delta E(L)/J_{ \perp} \gg 1$ is no longer fulfilled. This means the
Hilbert space is too severely truncated.

\subsection{Second order ground-state energies}

Table ~\ref{table2} displays second order ground-state energies for
a $12 \times 12$ system which are compared with QMC. The agreement is 
systematically better than in the first order case for all values of
$J_{\perp}$ studied. But the improvment is small. This is because 
as discussed above, the DMRG does not provide the full 1D spectrum. Only the 
states kept to form the reduced superblock are used in the perturbative 
expansion. When $ms_2=32$, this is merely $924$ states i.e., a tiny fraction of 
of the $2^{144}$ states which form the full Hilbert space of the 
$12 \times 12$ lattice. Another reason for this modest improvment is the
fact that the DMRG energies are variational. The high lying energies
which are used to generate second order terms are obtained with less
accuracy than the states kept in the first order. Indeed, this does not
mean that the matrix expansion presented above is not efficient. It has been
used in the simple case of the Mathieu equation for which the full spectrum
of unperturbed Hamiltonian is available \cite{MAT-moukouri}. The 
convergence of the matrix method
is quite impressive. Thus it seems that the best way to use the matrix 
Kato-Block expansion when the DMRG is used to obtain the unperturbed
spectrum is to restrict oneself to the first order and keep $ms_2$ as
large as possible. However, larger values of $ms_2$ are unpratical when
four blocks are used to form the superblock in the second step. For this
reason from now, only three blocks will be used to generate 
the superblock in the second step.  

\begin{table}
\begin{ruledtabular}
\begin{tabular}{ccccc}
 $J_{ \perp}$ & $ms_2=16$ & $ms_2=24$ & $ms_2=32$ & QMC  \\
\hline
 $0.00$ & -0.42848 & -0.42851 & -0.42851 & -0.42849(2) \\
 $0.05$ & -0.42901 & -0.42909 & -0.42910 & -0.42926(2) \\
 $0.10$ & -0.43063 & -0.43083 & -0.43094 & -0.43147(2) \\
 $0.15$ & -0.43322 & -0.43369 & -0.43396 & -0.43530(2) \\
 $0.20$ & -0.43661 & -0.43746 & -0.43797 & -0.44064(2) \\
 $0.25$ & -0.44055 & -0.44192 & -0.44271 & -0.44727(2) \\
\end{tabular}
\end{ruledtabular}
\caption{ DMRG ground state energies for $12 \times 12 $ lattices for 
second order DMRG compared with QMC for $ms_1=ms_2=16$, $24$ and $32$.}
\label{table2}
\end{table}

\section{First order results with three blocks in step 2}
 
When  three blocks are used, the superblock size in the second step is
divided by $ms_2$ relative to the case of four blocks. This significantly 
reduces the amount of required CPU for a given value of $ms_2$. But this
is not without problems. It was noted that \cite{white}, when three blocks 
are used, the coupling between blocks may incorrectly sets in leading to a 
poor performance of the method even if the truncation errors are small. The
remedy against this problem is to target more than one state so that
the interblock mixture is performed correctly. However, targeting many
superblock states lower the accuracy on the ground state. For this 
reason, only the ground state was targeted. The truncation errors were
in general smaller than $10^{-8}$ for $ms_2$ varying from $16$ to $96$
for different lattice size. But as said above, this does not give 
any indication about the accuracy of the second step of the method.
The QMC results are thus taken as the reference to gauge the DMRG
results.   

\subsection{Ground-state energies}

 For small sizes and weak couplings, the differences between the DMRG
 and QMC ground state energies  are very small. For instance for 
the $8 \times 9$ lattice shown in Table(~\ref{table3}), for $J_{\perp}=0.05$,
the difference between the two methods is only $0.00016$ for $ms_2=16$.
The two results are within QMC error when $ms_2$ is increased to $64$.  
 As expected, increasing the coupling tends to decrease the accuracy 
because the ratio $\delta E(L)/J_{ \perp}$ is reduced. Increasing the lattice
size has the same effect on this ratio because $\delta E(L)$ is smaller
for larger lattices for a fixed $ms_2$ 
(Table(~\ref{table3},~\ref{table4},~\ref{table5})). 
 One may note that by keeping a larger number of states than in the case
of four blocks, the accuracy has increased in all cases. 

\begin{table}
\begin{ruledtabular}
\begin{tabular}{ccccc}
 $J_{ \perp}$ & $ms_2=16$ & $ms_2=32$ & $ms_2=64$ & QMC  \\
\hline
 $0.00$ & -0.42187 & -0.42187 & -0.42187 & -0.42186(2) \\
 $0.05$ & -0.42239 & -0.42244 & -0.42247 & -0.42246(2) \\
 $0.10$ & -0.42402 & -0.42421 & -0.42439 & -0.42444(2) \\
 $0.15$ & -0.42670 & -0.42722 & -0.42762 & -0.42771(2) \\
 $0.20$ & -0.43032 & -0.43144 & -0.43219 & -0.43239(2) \\
 $0.25$ & -0.43470 & -0.43673 & -0.43799 & -0.43843(2) \\
\end{tabular}
\end{ruledtabular}
\caption{ DMRG ground state energies for $8 \times 9 $ lattices for
$ms_1=16$, and $ms_2=16$, $32$, and $64$ versus QMC.}
\label{table3}
\end{table}

\begin{table}
\begin{ruledtabular}
\begin{tabular}{ccccc}
 $J_{ \perp}$ & $ms_2=32$ & $ms_2=64$ & $ms_2=80$ & QMC  \\
\hline
 $0.00$ & -0.42851 & -0.42851 & -0.42851 & -0.42850(1)  \\
 $0.05$ & -0.42910 & -0.42918 & -0.42919 & -0.42922(1)  \\
 $0.10$ & -0.43094 & -0.43124 & -0.43131 & -0.43150(1)  \\
 $0.15$ & -0.43396 & -0.43468 & -0.43483 & -0.43537(1)  \\
 $0.20$ & -0.43796 & -0.43928 & -0.43956 & -0.44075(1)  \\
 $0.25$ & -0.44268 & -0.44476 & -0.44521 & -0.44744(1)  \\
\end{tabular}
\end{ruledtabular}
\caption{ DMRG ground state energies for  $12 \times 13$
 lattices for $ms_1=ms_2=32$, $ms_2=64$ and $ms_2=80$ ($ms_1=64$
for both) versus QMC.}
\label{table4}
\end{table}

\begin{table}
\begin{ruledtabular}
\begin{tabular}{ccc}
 $L\times L+1$ & DMRG &  QMC  \\
\hline
 $8 \times   9$ & -0.42440 & -0.42444(2) \\
 $12 \times 13$ & -0.43124 & -0.43150(2) \\
 $16 \times 17$ & -0.43481 & -0.43529(1) \\
\end{tabular}
\end{ruledtabular}
\caption{ DMRG ground state energies for various lattices for
$ms_2=80$ and $J_{\perp}=0.1$.}
\label{table5}
\end{table}

\subsection{Ground-state correlation functions}

It is not possible to keep track of all spin-spin correlations
when large systems are studied because of CPU and memory limitations.
The behavior of spin-spin correlations is thus studied along one
chain in the direction parallel to the chains and one chain in the
direction perpendicular to the chains. These correlation functions
are respectively given below:

\begin{eqnarray}
C_{\parallel}(i,l,r)=\frac{1}{3}\langle{\bf S}_{i,l}.{\bf S}_{i+r,l}\rangle,\\
C_{\perp}(i,l,r)=\frac{1}{3}\langle{\bf S}_{i,l}.{\bf S}_{i,l+r}\rangle
\end{eqnarray}

It is particularly difficult to obtain the large $r$ behavior 
 of the correlation functions because of a number of
factors that complicate such an analysis. At the level of
a single chain, the long distance behavior of $C(i,r)$ is already 
complicated by logarithmic corrections. Although highly accurate
data can be obtained in 1D from QMC \cite{sandvik3} or DMRG \cite{hallberg},
 the two studies disagree on the exact form of the logarithmic 
corrections. Furthermore when open boundary
conditions (OBC) are used instead of periodic, the spin-spin correlation
functions show strong odd-even alternations \cite{white, caron}.
This is because the ground state may
be regarded as a resonant state between a state with strong bonds on even
links and weak bonds on odd links, and a state with weak bonds on even links
and strong bonds on odd links.
Another difficulty with OBC is that the translational invariance of the chain
is broken, and the value of $C_{\parallel}(i,l,r)$ depends on the position
of the site chosen as the origin on the lattice. It was shown \cite{caron}
that the closer the origin is to the edge of the lattice, the
higher are the spurious effects introduced by the OBC.
All these facts render the direct detection of long range order in the
transverse direction, for which the spin-spin correlations are very
small, impossible to achieve with the present calculation for which the
magnitude of $C_{\perp}(i,l,r)$ for large $r$ is 
 close to the accuracy on
the eigenvalues during each iteration. An alternative way is to
look at the  $C_{\parallel}(i,l,r)$, because the existence of long range
order in the longitudinal direction is an indication that the order
is two-dimensional.

\begin{table}
\begin{ruledtabular}
\begin{tabular}{ccccc}

 $l$ & DMRG (l) & QMC (l) & DMRG (t) & QMC (t) \\
\hline

 1 & -0.14595 & -0.14931(1) & -0.02047 & -0.02209(1) \\
 2 &  0.06072 &  0.05904(1) &  0.00561 &  0.00525(1) \\
 3 & -0.04799 & -0.05173(1) & -0.00191 & -0.00164 \\
 4 &  0.03340 &  0.03537(1) &  0.00066 &  0.00055 \\
 5 &          &             & -0.00023 & -0.00019 \\

\end{tabular}
\end{ruledtabular}
\caption{DMRG versus QMC longitudinal (l) ${\bar C}_{\parallel}(7,7,r)$ and
transverse ${\bar C}_{\perp}(7,7,r)$ spin-spin correlations
for a $12 \times 13$ lattice for $J_{ \perp}=0.1$, $J_{d}=0$. }
\label{table6}
\end{table}

 In order to observe the correct long-range behavior, one must first reduce
the influence of the spurious effects generated by the application 
of the OBC. Furthemore to simplify the analysis, the eventual logarithmic 
corrections will not be considered here. In order to avoid the odd-even 
alternation, $C_{\parallel}(i,l,r)$ and $C_{\perp}(i,l,r)$ were averaged 
in the period of these alternations.  This was done by computing 
$\langle{\bf S}_{i,l} {\bf S}_{i+r,l}\rangle$ at two different origins. 
The spin 
${\bf S}_{il}$ is taken as the origin of a strong link or as the origin of a 
weak link. The actual correlation function is then

\begin{eqnarray}
{\bar C}_{\parallel}(i,l,r) =0.5(C_{\parallel}(i,l,r)+
 C_{\parallel}(i+1,l,r)).    
\end{eqnarray}

And for $C_{\perp}(i,l,r)$,

\begin{eqnarray}
{\bar C}_{\perp}(i,l,r) =0.5(C_{\perp}(i,l,r)+
 C_{\perp}(i,l-1,r)).    
\end{eqnarray}

\begin{table}
\begin{ruledtabular}
\begin{tabular}{ccccc}

 $l$ & DMRG (l) & QMC (l) & DMRG (t) & QMC (t) \\
\hline

 1 & -0.14640 & -0.14619(1) & -0.02116 & -0.02533(1) \\
 2 &  0.06059 &  0.06130(1) &  0.00726 &  0.00854(1) \\
 3 & -0.04875 & -0.04988(1) & -0.00320 & -0.00399 \\
 4 &  0.03422 &  0.03537(1) &  0.00147 &  0.00201 \\
 5 & -0.02866 & -0.02990(1) & -0.00078 & -0.00105 \\
 6 &  0.02251 &  0.02363(1) &  0.00030 &  0.00056 \\
 7 &          &             & -0.00013 & -0.00030 \\
 8 &          &             &  0.00006 &  0.00015 \\

\end{tabular}
\end{ruledtabular}
\caption{DMRG versus QMC longitudinal (l) ${\bar C}_{\parallel}(9,9,r)$ and
transverse ${\bar C}_{\perp}(9,9,r)$ spin-spin correlations
for a $16 \times 17$ lattice for $J_{ \perp}=0.1$, $J_{d}=0$. }
\label{table7}
\end{table}

The averaged correlations ${\bar C}_{\parallel}(i,l,r)$ and 
${\bar C}_{\perp}(i,l,r)$ are shown in Tables~\ref{table6}, ~\ref{table7} and
~\ref{table8} for, respectively, $12 \times 13$, $16 \times 17$ 
and $32 \times 33$
lattices. The origins $(i,l)$ of the correlation functions were chosen
at the middle of the chain in order to minimize the end effects.
 $(i,l)$ was equal to $(7,7)$, $(9,9)$ and $(17,17)$ respectively for
the $12 \times 13$, $16 \times 17$ and $32 \times 33$ lattices.
For the $16 \times 17$ lattice, $ms_1=128$ states were kept
during the first DMRG step and $ms_2=64$ states were kept during the 
second DMRG step. The comparison with QMC is quite good in the
longitudinal direction but less  good in the transverse direction when
the lattice size gets large.
 For the $32 \times 33$ lattice, $ms_1$ and $ms_2$ were respectively
increased to $160$ and $80$. As for the case of the
$16 \times 17$ lattice the agreement was quite good for 
${\bar C}_{\parallel}(i,l,r)$ and less good for ${\bar C}_{\perp}(i,l,r)$.
The reasons for the differences are not easy to analyze. Although
very small truncation errors $p_m$ (for instance, $p_m < 1\times 10^{-7}$ 
for $ms_1=128$ and $ms_2=64)$) are obtained in the DMRG, there is no 
obvious relation between these truncation errors and the errors on the 
measurements. Furthermore, the effects of higher order terms in the 
perturbation series have not be analyzed for the case of three blocks.
Since more states are kept when three blocks are used, the contribution
of second order terms is likely larger than the one found above for four
blocks.

\begin{table}
\begin{ruledtabular}
\begin{tabular}{ccccc}

 $l$ & DMRG (l) & QMC (l) & DMRG (t) & QMC (t) \\

\hline
 1 & -0.14694 & -0.14636(3) & -0.01846 &  -0.02952(2) \\
 2 &  0.06042 &  0.06151(3) &  0.00969 &   0.01465(2) \\
 3 & -0.04908 & -0.05066(2) & -0.00623 &  -0.01057(3) \\
 4 &  0.03402 &  0.03640(2) &  0.00416 &   0.00821(1) \\
 5 & -0.02949 & -0.03229(3) & -0.00281 &  -0.00662(1) \\
 6 &  0.02366 &  0.02682(3) &  0.00190 &   0.00545(1) \\
 7 & -0.02108 & -0.02450(2) & -0.00128 &  -0.00453(2) \\
 8 &  0.01820 &  0.02163(3) &  0.00086 &   0.00379(2) \\
 9 & -0.01643 & -0.01990(2) & -0.00059 &  -0.00321(2) \\
10 &  0.01474 &  0.01806(2) &  0.00040 &   0.00270(2) \\
11 & -0.01327 & -0.01646(2) & -0.00028 &  -0.00228(2) \\
12 &  0.01202 &  0.01508(2) &  0.00018 &   0.00194(2) \\
13 & -0.01045 & -0.01309(1) & -0.00012 &  -0.00164(2) \\
14 &  0.00914 &  0.01164(2) &  0.00008 &   0.00137(2) \\
15 &          &             & -0.00005 &  -0.00115(1) \\
16 &          &             &  0.00003 &   0.00088(1) \\

\end{tabular}
\end{ruledtabular}
\caption{DMRG versus QMC longitudinal (l) ${\bar C}_{\parallel}(17,17,r)$, and
transverse (t) ${\bar C}_{\perp}(17,17,r)$ spin-spin correlations
for a $32 \times 33$ lattice for $J_{ \perp}=0.1$, $J_{d}=0$. }
\label{table8}
\end{table}

\section{Ground-state properties in presence of frustration}
\label{frustration}

The DMRG method has shown an overall good agreement with QMC
for weak couplings and not too large sizes. The method is well
controlled and can systematically be improved by increasing
$ms_1$ and $ms_2$. The advantage of the DMRG over QMC is that
it is very flexible and can be applied to frustrated systems.
A situation where the QMC is known to fail. In this section
a diagonal $J_d$ exchange coupling is included. It has the effect
of introducing a competition between interchain AFM correlations along 
the rows and AFM correlations along the diagonals.

\subsection{Ground-state energies}

\begin{figure}
\includegraphics[width=3. in]{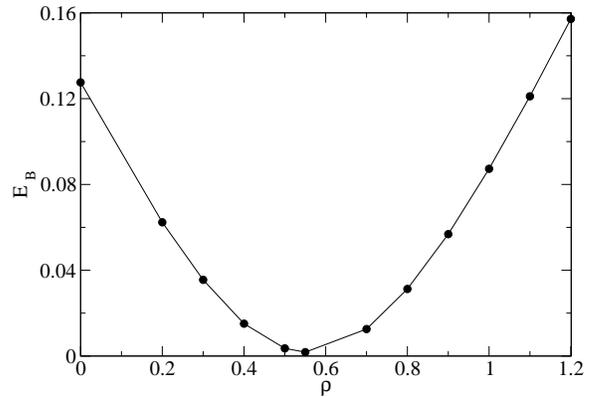}
\caption{The binding energy $E_B$ with respect to single chain 
for a $32 \times 33$ lattice as a function of $\rho=J_{d}/J_{ \perp}$,
 $J_{ \perp}=0.1$.} 
\label{egm64l32}
\end{figure}

Although the result on the ground state energy can not provide 
information about a possible long-range order, it can be helpful
to see if the perturbation is relevant or not. Fig. ~\ref{egm64l32}
shows the binding energy per chain $E_B=E_0(L)-E_0(L \times (L+1))/(L+1)$, 
where
$E_0 (L)$ is the ground state energy for a single chain and 
$E_0(L \times (L+1))$ is the ground state energy for an 
$L \times (L+1)$ lattice. $J_{ \perp}$ is set to
$0.1$. $E_B$ first decreases as $J_{d}/J_{ \perp}$ is increased.
It reaches a minimum at $ J_{d} \approx 0.5 J_{ \perp}$. At the
minimum point, the binding energy nearly vanishes, $E_B \approx 0.0015$ which
is roughly two orders of magnitude smaller than its value 
for $J_{d}=0$. As 
$J_{d}/J_{ \perp}$ is further increased, $E_B$ starts increasing
sharply. This behavior suggests the existence of three regimes for
for the action of small perturbations on the single chain, two stable
phases separated by a transition region. The first regime, which
occurs  when $J_{d} \alt 0.5 J_{ \perp}$, is a N\'eel state as 
is already known from QMC studies \cite{sandvik2}. This will be confirmed 
below by the analysis of the DMRG correlation functions. The second regime
is when $J_{d} \approx 0.5 J_{ \perp}$. The perturbation seems  
to be irrelevant,  $J_{ \perp}$ and $J_{d}$ cancel each other
so that there is almost no gain in energy by applying the two perturbations
simultaneously. In the third regime, when $J_{d} \agt 0.5 J_{ \perp}$,
the ground state is also magnetic with a collinear order, an alternate 
arrangement of transverse up and down ferromagnetic chains 
(see Fig.~\ref{gstate}).

\begin{figure}
\includegraphics[width=3. in]{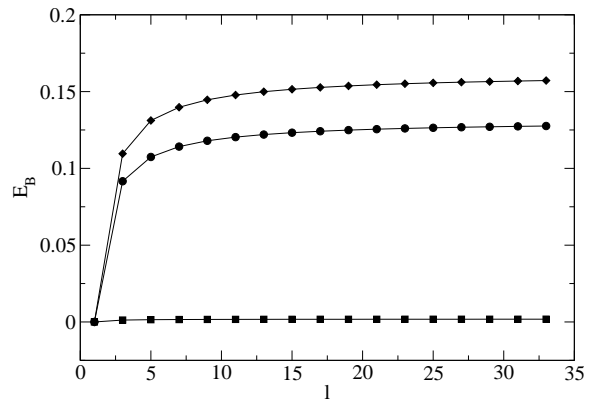}
\caption{The binding energy $E_B$ with respect to single chain 
for a $32 \times l$ lattice as a function of $l$ for 
$J_{d}/J_{ \perp}=0$ (circles), $J_{d}/J_{ \perp}=0.5$ (squares),
$J_{d}/J_{ \perp}=1.2$ (diamonds), $J_{ \perp}=0.1$.}
\label{egm64l3-33}
\end{figure}

The above analysis is further supported by observing the evolution
of the binding energy $E_B(L \times l)$ as a function of 
the number of chains
in the lattice (Fig. ~\ref{egm64l3-33}). It clearly shows that when
$J_{d} \approx 0.5 J_{ \perp}$, the binding energy is nearly independent
of the number of chains and remains very close to that of the single
chain. Hence it seems that at the point 
$J_{d} \approx 0.5 J_{ \perp}$, the ground state is made of
independent chains as for $J_{ \perp}=J_{d}=0$.
This behavior is analogous to the domino model 
studied by Villain and coworkers \cite{villain},  where a disordered
ground state, made of independent chains for a particular
value of the transverse coupling, was found.

\subsection{Ground-state correlation functions}

\begin{figure}
\includegraphics[width=3. in]{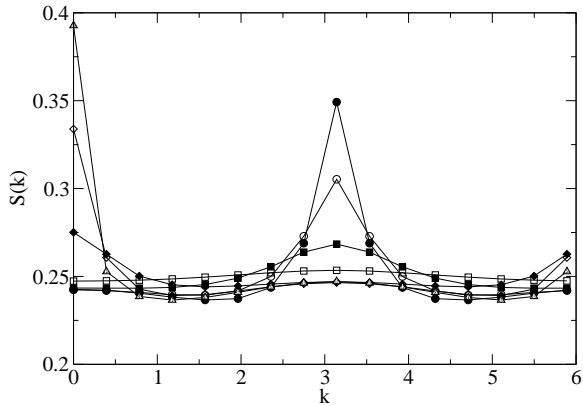}
\caption{The ground state transverse structure factor for a $32 \times 33$ 
lattice for
 $J_{d}/J_{ \perp}=0$ (filled circles), $0.2$ (open circles), 
 $0.4$ (filled squares), $0.55$ (open squares), $0.8$ (filled diamonds),
 $1.0$ (open diamonds), $1.2$ (open triangles).} 
\label{strf}
\end{figure}

The behavior of the correlation functions is consistent with the
existence of the three regimes found for the ground state energy.
As expected, spin-spin correlations along the chains remain
antiferromagnetic. The change of regimes  will be detected by analyzing
spin-spin correlations along the transverse direction. Fig.~\ref{strf}
 shows the transverse magnetic structure factor $S_{\perp}(k_{\perp})$,

\begin{eqnarray}
S_{\perp}(k_{\perp})= \sum_{k_{\perp}=1}^{L/2} {\bar C}_{\perp}(17,17,r) 
 \cos k_{\perp}r
\label{strfac}
\end{eqnarray}

\noindent where $k_{\perp}$ is a wave number in the transverse direction. 
It also shows the three regimes
discussed above. When  $J_{d} \alt 0.5 J_{ \perp}$, 
$S_{\perp}(k_{\perp})$ has
a maximum at $k_{\perp}=\pi$. The spin-spin correlations along the tranverse
direction are AFM as for the longitudinal direction. For 
$J_{d} \approx 0.5 J_{ \perp}$, $S_{\perp}$ is structureless,
a fact which is consistent with disconnected chains.  When 
$J_{d} \agt 0.5 J_{ \perp}$, $S_{\perp}$ has a maximum at
$k_{\perp}=0$, and the correlations in the transverse direction are 
now ferromagnetic.
This is the collinear magnetic state shown in Fig.~\ref{gstate}.

\begin{figure}
\includegraphics[width=3. in]{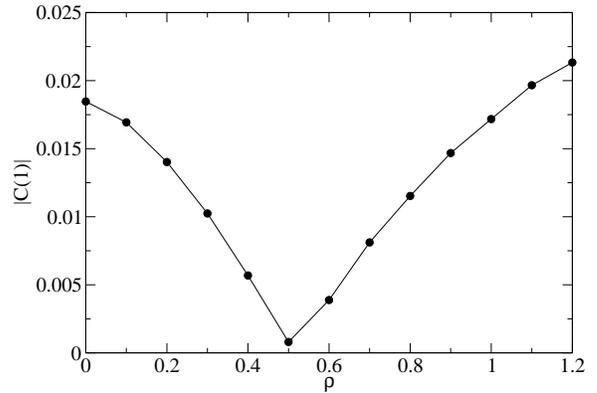}
\caption{The bond-strength $C(1)={\bar C}_{\perp}(17,17,1)$ as a function
of $\rho=J_{d}/J_{ \perp}$, $J_{ \perp}=0.1$.}
\label{bond}
\end{figure}

The bond-strength ${\bar C}_{\perp}(17,17,1)$, computed in a $32 \times 33$ 
lattice is shown in Fig.~\ref{bond}.
 It also shows that the chains seem to be  disconnected when 
$J_{d} \approx 0.5 J_{ \perp}$. Starting from $J_{d}=0$,
for which  ${\bar C}_{\perp}(17,17,1)= -0.01846$, its absolute value 
first slowly
decreases. Then, when $J_{d}$ is in the vicinity of $J_{d_C}$,
the absolute value of  ${\bar C}_{\perp}(17,17,1)$ sharply decreases and
become very small; ${\bar C}_{\perp}(17,17,1)= -0.000799$ when 
$J_{d}=0.5 J_{ \perp}$. As soon as $J_{d}$ exceeds 
$J_{d_C}$, ${\bar C}_{\perp}(17,17,1)$ becomes ferromagnetic and starts to
increase sharply. It later saturates when one is far enough from
the critical point. 

\section{Long-range order in the ground state}
\label{longrange}

The analysis made in the preceeding section indicates regions of
dominant N\'eel or collinear spin-spin correlations or of a possibly 
disordered ground state at the transition point. But it does
not tell if long range order is truly established. For this, it
is necessary to look at the long-range behavior of the correlation
functions. 

The spurious effects due to the breaking of the translational
symmetry, a consequence of the OBC, may be reduced by using a 
filter which smooths the
action of the sites near the edges. In the results shown below in
Fig ~\ref{spincorl1},~\ref{spincorl2},~\ref{spincorl3}, 
this was done as follows: ${\bar C}_{\parallel}(i,l,r)$  was first 
examined for a single chain for which the long distance behavior
is known. Roughly, ${\bar C}_{\parallel}(i,l,r) \propto 1/r$ if logarithmic
corrections are neglected. It was found that if the origin is taken
at the middle of the chain, the $1/r$ behavior is roughly satisfied
for $d \alt r \alt L/2-d$ with $d \approx 8$.  The second inequality 
is due to edge effects.
As a consequence, relatively large values of $L$ are necessary in order
to observe the long range behavior, and lattices of up to $64 \times 65$
were studied. The problem with such large lattices
is that the energy width $\delta E(L)$ shrinks with increasing $L$ 
and the condition 
$\Delta_{\sigma}(L) \simeq J_{\perp,d} \ll \delta E(L)$ may not be 
fulfilled.  For $L=64$, $ms_1=192$ and $ms_2=80$ states were kept.
 For these values,  $\delta E(L=64) \approx 0.5$, which
means $\delta E(L=64)/J_{\perp,d} \approx 5$ provided that 
$J_{\perp,d} \alt 0.16$.

\begin{figure}
\includegraphics[width=3. in]{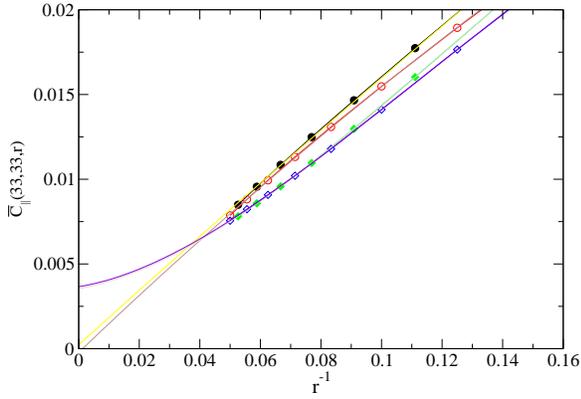}
\caption{The ground state correlation function 
${\bar C}_{\parallel}(33,33,r)$ for the $64 \times 65$
lattice for $J_{ \perp}=0$ (circles) and $J_{ \perp}=0.16$ (squares).
$ J_{d}=0$ in both cases. The filled and open symbols correspond to 
odd and even distances respectively.} 
\label{spincorl1}
\end{figure}

The first question which needs to be addressed is to know whether the
DMRG can detect an eventual long range order. Comparisons with QMC for
$L=32$ show that, the DMRG correlation in the transverse direction
decays faster. This effect is expected to be larger on longer chains.
 But, despite this shortcoming of the method, one can still detect
possible occurrence of long range order. If one considers the central
chain in the 2D lattice, ${\bar C}_{\parallel}(L/2+1,L/2+1,r)$ is modified
from that of an isolated chain because of the effective magnetic field
created on it by the rest of lattice. Although this effective field is
somewhat underevaluated by the DMRG because the transverse correlations
are underevaluated, it can still be strong enough to lead to an ordered
phase. This interpretation is related to the chain mean-field 
approach \cite{scalapino};
the essential point is that, here, no assumption about long-range order is 
made {\it a priori}. From this, one can see that if the DMRG method leads to
a finite order parameter, it is necessarily genuine. Fig.~\ref{spincorl1}
 compares for the correlation function ${\bar C}_{\parallel}(33,33,r)$
for $J_{ \perp}=0$, $J_{d}=0$ with  $J_{ \perp}=0.16$, 
$J_{d}=0$. In the first case when both transverse couplings are
absent,  ${\bar C}_{\parallel}(33,33,r) \propto 1/r $. The DMRG data
still show an odd-even alternation, so fits must be perfomed for
odd and even distances separately. The best least square fits to the
data gave ${\bar C}_{\parallel}(33,33,r \rightarrow \infty) \approx 0.0001$.
 This is consistent with an absence of a long-range order for an
isolated chain. But in the case $J_{ \perp}=0.16$ and $J_{d}=0$,
a fit to the data shows that ${\bar C}_{\parallel}(33,33,r)$ tends to
${\bar C}_{\parallel}(33,33,r \rightarrow \infty) \approx 0.0036$  
The existence of long-range N\'eel order for $J_{ \perp}=0.16$ is consistent
with previous studies \cite{sandvik2,affleck}.  
 Adding $J_{ \perp}$ alone seems to lead to long-range order. This 
has been recently shown in Ref.\cite{sandvik2} where values
of $J_{ \perp}$ down to $0.02$ were investigated.  It is of course
impossible to show from a numerical investigation whether any small value of
$J_{ \perp}$ will lead to an ordered state  or there may be a disordered
state for very small values of $J_{ \perp}$. In view of current
numerical results, the former hypothesis is more convincing.

\begin{figure}
\includegraphics[width=3. in]{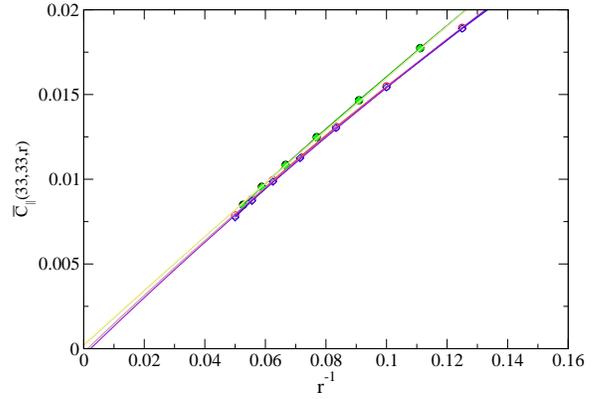}
\caption{The ground state correlation function 
${\bar C}_{\parallel}(33,33,r)$ for the $64 \times 65$
lattice for $J_{ \perp}=0.$, $ J_{d}=0.$  (circles) and 
$J_{ \perp}=0.16$, $ J_{d}=0.08$  (squares).
 The filled and open symbols correspond to
odd and even distances respectively.}

\label{spincorl2}
\end{figure}

\begin{figure}
\includegraphics[width=3. in]{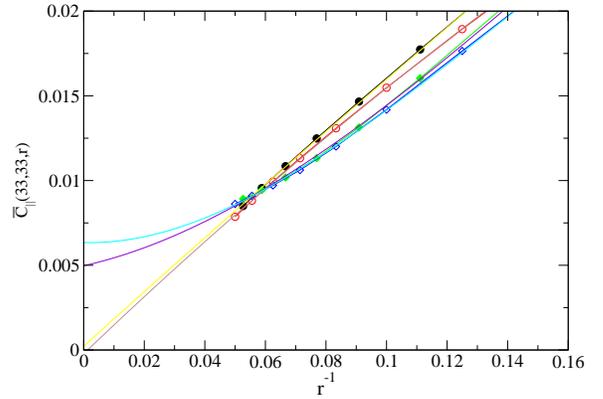}

\caption{The ground state correlation function
${\bar C}_{\parallel}(33,33,r)$ for the $64 \times 65$
lattice for $J_{ \perp}=0.$, $ J_{d}=0.$  (circles) and
$J_{ \perp}=0.16$, $ J_{d}=0.16$  (squares).
 The filled and open symbols correspond to
odd and even distances respectively.}
\label{spincorl3}
\end{figure}

The above discussion suggests that a frustration $J_{d}$ must
be added in order to thwart the N\'eel state which results from the
action of $J_{ \perp}$. $J_{ \perp}$ will now be set to $0.16$ and
$J_{d}$ varied. For $J_{d} = 0.08$, the value at which
the analysis of smaller chains suggested that the ground state is
made of disconnected chains, ${\bar C}_{\parallel}(33,33,r)$ is 
compared to the same quantity for a single chain in Fig.~\ref{spincorl2}.
Fits to the data show that the behavior of ${\bar C}_{\parallel}(33,33,r)$ is
quite similar to that of a single chain. Clearly, for these values of the
transverse couplings there is no long-range order in
the ground state, and ${\bar C}_{\parallel}(33,33,r)$ seems to indicate that 
the ground state is made of a set of independent chains. It is important 
to emphasize that this result does not mean that the finite temperature 
effects are also trivial.
 The present situation could be similar to the so-called domino model
first introduced by Andre \cite{andre} and later studied by Villain
and coworkers \cite{villain} or to the crossed-chains quantum spin
models \cite{singh}. In the domino model, it was found that
the ground state was made of disconnected chains but there was a 
long-range order at finite temperature. Indeed, the Mermin-Wagner
theorem prohibits long-range order at finite temperatures for the
2D Heisenberg model. The finite temperature behavior in this case 
will thus be different.

 The disconnected chain ground state is in contradiction with a recent
study by Nersesyan and Tsvelik \cite{tsvelik}. 
These authors argued, using bosonization,
that when $J_{d}/J_{ \perp} =0.5$, only the staggered part of the
interchain part of the Hamiltonian vanishes. There remains a uniform
part which is relevant and leads to two-dimensional spin liquid with
a spin gap, 
$\Delta_{\sigma} \propto \exp (- \frac{\pi v_{\sigma}}{2J_{ \perp}})$,
 where $v_{\sigma}$ is the spin velocity. The low energy excitations are
argued to be unconfined spinons. The apparent contradiction between
this conclusion and the numerical data above could be that the binding energies
of the 2D spin liquid are very small, indeed $J_{ \perp} =0.1$ corresponds
to $\Delta_{\sigma} \approx 1.0 \times 10^{-11}$. 
Such a small energy can obviously
not be detected by a numerical method. A way to avoid this small
energy scale is to raise $J_{ \perp}$. This possibility is currently
being investigated.

 Finally, the collinear magnetic long range order is also confirmed by the
analysis of ${\bar C}_{\parallel}(33,33,r)$. In Fig.~\ref{spincorl3},
it is shown that for $J_{d}=0.1$, ${\bar C}_{\parallel}(33,33,r)$
converges even faster than for the N\'eel state above. The value
of the extrapolated correlation is 
${\bar C}_{\parallel}(33,33,r \rightarrow \infty) \approx 0.0056$.

\section{Conclusions}
\label{conclusion}

In this paper, a new renormalization group method for weakly 
coupled chains was presented. It is based on solving numerically 
the model Hamiltonian in two 1D steps using the DMRG. 
During the first step, a low energy Hamiltonian for a single 
chain is obtained using the 1D DMRG. The original problem is then
formulated as a perturbative expansion around the DMRG low energy 
Hamiltonian  obtained during the first step. This perturbative
expansion is a 1D problem which can also be solved by the DMRG.

The first  and second order approximations 
were studied for weakly coupled Heisenberg chains with and without
frustration. The results were compared to the QMC and showed
good agreement for small systems and small transverse couplings. 
It was shown that, starting from the disordered
1D chain, the method can predict long-range order when it exists,
a test generally failed by conventional perturbative methods.
Calculations performed in the presence of frustration indicate an
absence of a genuinely 2D spin liquid state. Instead, the frustration
drives the N\'eel ground state to a collinear magnetic state.
At the transition point, both ground-state energy and spin-spin
correlation functions show a disordered ground state. The precise nature
of this disordered ground state is currently under investigation.

The above results are very encouraging and indicate that the DMRG may
become a very useful tool for the study of highly anisotropic 2D
systems in the future. The method is only in its early stages, and some
important improvements of the method are currently underway. These are
 the investigation of the role of cluster corrections, i.e., the starting 
point in the first step will be two-leg or three-leg ladders instead of a 
single chain; the use of exact diagonalization during the first step instead
of DMRG. These
improvements are likely to lead to better results for spin-spin correlations
in the transverse direction. Extensions of the method to thermodynamic 
 spin systems or fermionic models will also be made in the near future.

\begin{acknowledgments}
 I am very grateful to A. Sandvik for sharing his QMC data and for numerous
 helpful exchanges during the course of this work. I wish to thank 
J.V. Alvarez for useful discussions. I also thank J.W. Allen and P.
McRobbie for reading the manuscript. 
\end{acknowledgments}

\end{document}